# MnP films with desired magnetic, magnetocaloric and thermoelectric properties for a perspective magneto-thermo-electric cooling device

C.M. Hung[1], R.P. Madhogaria[1], B. Muchharla[1], E.M. Clements[1], A.T. Duong[2,3,*], R. Das[2], P.T. Huy[2], S.L. Cho[4], S. Witanachchi[1], H. Srikanth[1,*], and M.H. Phan[1,*]

[1] Department of Physics, University of South Florida, Tampa, Florida 33620, USA

[2] Faculty of Materials Science and Engineering, Phenikaa University, Yen Nghia, Ha Dong, Hanoi 12116, Vietnam

[3] Phenikaa Research and Technology Institute (PRATI), A&A Green Phoenix Group, 167 Hoang Ngan, Hanoi 13313, Vietnam

[4] Department of Physics and Energy Harvest-Storage Research Center, University of Ulsan, Ulsan 680-749, Republic of Korea

**A perspective magneto-thermo-electric cooling device (MTECD) comprising a central magnetocaloric (MC) material (e.g., Gd) sandwiched by two thermoelectric (TE) materials (e.g., MnP) is proposed. The presence of the TE materials in the MTECD guides the heat flow direction and enhances heat pulsation. In this case, the usage of a ferromagnetic TE material that combines large TE with small MC properties within a similar temperature region can enhance the magnetic flux density and heat exchange efficiency. Here, we show that MnP nanorod-structured films with desired magnetic, MC and TE properties are very promising for use in MTECDs. The films were grown on Si substrates at 300, 400 and 500°C using molecular beam epitaxy. The 400 ºC sample shows a desired TE and MC combination. A large power factor of 24.06 μW m$^{-1}$ K$^{-2}$ is achieved at room temperature. In this temperature region, the film exhibits a small MC effect (-$\Delta S_M$ ~0.64 J/kg K and $\Delta T_{ad}$ ~0.3 K at $\mu_0 H$ = 2 T) but ferromagnetism that gives rise to the enhanced MC effect of the central MC material. These properties could enable the MTECD to operate at high frequency.**

*Corresponding authors: tuan.duonganh@phenikaa-uni.edu.vn (A.T.D); sharihar@usf.edu (H.S); phanm@usf.edu (M.H.P)

## 1. Introduction

Refrigeration is vital to our daily life. Current refrigerators operate based on conventional gas compression/expansion techniques. However, these techniques have reached the upper limit of cooling efficiency and utilize greenhouse gases (e.g. hydrofluorocarbons), which trigger global climate change when escaped into the atmosphere. Alternative cooling techniques based on caloric effects without greenhouse gasses (e.g., magnetocalorics[1], electrocalorics[2], elastocalorics[3], spincalorics[4]) have been extensively explored over the past decades, owing to their potentially high efficiency, compactness, and environmental friendliness[5-8]. Among them, magnetic refrigeration based on the magnetocaloric effect (MCE) holds an enormous promise[9], since the giant MCEs, which yield the high cooling efficiency of the device, have been discovered in a wide range of magnetic materials (e.g. $Gd_5Si_2Ge_2$[10], La–Fe–Si–Mn–H[11], Mn–Fe–P–Si[12], Ni-Mn-In[13], and high entropy alloy[14]). However, these coolers are restricted to operate at low frequency (a few tens of Hz), as a direct result of relatively poor heat transfer between the magnetic refrigerant (the magnetocaloric material used) and surrounding materials towards the heat sink[15]. Numerous efforts have been therefore devoted to the design of more energy-efficient cooling systems in which film- or wire-shaped magnetocaloric materials can be assembled into a laminate structure or combined with other materials with high thermal conductivity.[16-19]

In this regard, combining one magnetocaloric (MC) material with two thermoelectric (TE) materials in a sandwich structure to promote fast heat exchange/transfer from the magnetocaloric material to the heat sink, as proposed by Kitanovski and Egolf[15], represents a promising approach for the design of an energy-efficient cooling device. In such a device, the MC and TE materials are required to exhibit large magnetocaloric and thermoelectric effects over a similar temperature region (ideally around room temperature), respectively. It has been suggested that if the TE materials are ferromagnetic, the magnetic flux density in the MC material subjected to an external magnetic field can be enhanced, giving rise to the large cooling efficiency.[15] However, the applied magnetic field may rise the temperature of the TE material due to the MCE, hampering its heat transfer to the heat sink.[20] To overcome this challenge, it is essential to design a TE material that possesses combined ferromagnetic, *large* TE and *small* MC properties around room temperature.

In this paper, we propose a perspective magneto-thermo-electric cooling device (MTECD) that comprises one magnetocaloric and two thermoelectric materials, in which the thermoelectric materials are required to possess large TE and small MC responses simultaneously. Magnetic, magnetocaloric and thermoelectric measurements indicate that high quality MnP films with a combination of these properties are attractive candidate materials for use in energy-efficient cooling devices.

## 2. Method and Experiment

Fig. 1(a) illustrates the design of a MTECD, in which one magnetocaloric material (MCM, red) is sandwiched between two thermoelectric materials (TEMs, blue) and separated by two thin insulating materials (spacers, yellow). The outermost part of the device is equipped with the micro-channels (gray), in order to increase the heat-exchange surface area. The highlight of the device is that the TE materials are soft ferromagnets (e.g., the MnP film as reported in this study). The working principle of the designed device is described in Fig. 1(b). When an external magnetic field is applied parallel to the plane of the film, it magnetizes the MCM, resulting in the increase of temperature of the material. At this moment, a DC current is applied which passes through the upper TEM, creating an active thermoelectric cooling device (TCD). Meanwhile, the bottom TEM does not allow the heat to flow because no current is running through the bottom TEM. The heat flowing through the upper TEM reaches the micro-channels through which it is released to the heat sink. During this process, the two TEMs act as an extra magnetic field source which increases the total magnetic flux density of the MCM. When the magnetic field is off, the temperature of the MCM is reduced, and the DC current now passes through the lower TEM, which absorbs the heat from the heat source and transmits it to the MCM. The heating/cooling cycle is repeated. It is worth noting that the thickness of both MCM and TEM layers should be thin so that the heat transfer can occur easily in a desired direction. In addition to the above criteria, the MCM should possess high thermal conductivity ($\kappa$), while the TEMs should have low $\kappa$ to ensure a temperature gradient. A potential MCM for use in this type of device is, for example, gadolinium (Gd)[21] exhibiting the large MCE ($\Delta T_{ad}$ = 5.8 K for 2 T) with negligible magnetic/thermal hysteresis due to its second order magnetic transition (SOMT), as well as its high thermal conductivity at room temperature. $Pr_2F_{17}$[22] with a large working temperature range (125 K) around room temperature can also be a promising candidate. Other materials of potential interest may be found in the work of Ucar *et al.*[23] For the TE materials operating in the room temperature regime, we show below that our MnP nanorod-structured films with combined ferromagnetic, large TE and small MC responses are a very promising candidate for use in MTECDs.

MnP films were grown on Si(1 0 0) substrates using molecular beam epitaxy. The films were grown at 300, 400, and 500°C (for short, we denoted them as the 300, 400, and 500°C samples/films) and further details of the films' growth can be found in Duong *et al.*[24] The crystal phase characterization of the MnP thin films was carried out using Bruker AXS powder X-ray diffractometer (XRD) with Cu-K$_\alpha$ radiation at room temperature. Field emission scanning electron microscopy (FE-SEM) was performed to observe microstructural variation in the films grown at different temperatures. Magnetic properties were measured by a Vibrating Sample Magnetometer (VSM) equipped within the Physical Property Measurement System (PPMS) from Quantum Design.

Magnetization versus temperature was measured from $T$ = 150 - 370K for $\mu_0 H$ = 0.01- 2 T using field-cooled warming measurement protocols, and magnetization versus magnetic field were performed at $\mu_0 H$ = 0 - 5 $T$ for temperatures $T$ = 200 - 370 K. The electric transport properties were characterized using a standard four-probe measurement system. Thermoelectric properties of the MnP films were measured in the temperature range of 20- 400 K.

**3. Results and Discussion**

The XRD patterns of MnP thin films grown at 300, 400 and 500°C are shown in Fig. S1(a) (Supplementary Information). The samples grown at different temperatures exhibited similar orthorhombic MnP peaks. The FE-SEM images of 300, 400 and 500°C grown films are shown in Fig. S1(b), (c) and (d), respectively. The SEM images depict the out-of-plane growth of nanorods of varying sizes (20 – 100 nm) depending on the growth temperatures. As shown in Fig. S1(b-d), the size of nanorods increases with increase in the growth temperature, resulting in larger strain.[24] Moreover, the MnP crystals grown at 500 °C were shaped like nanoclusters rather than nanorods, implying that the crystallinity of the film grown at 400°C is optimal. The inset of Fig. S1(c) displays a top-view TEM image of the 400°C sample, which shows a clear structure of the MnP nanorods.

The in-plane temperature-dependent magnetization (field-cooled warming) *M-T* curves, for $T$ > 150 K and $\mu_0 H$ = 0.1 T, are shown in Fig. 2(a) for all samples grown. From Fig. 2(a), all the three samples display a clear ferromagnetic to paramagnetic (FM-PM) transition around room temperature. The magnetization tends to increase for the films grown at higher temperatures (Fig. 2(a)), owing to the increased size of the nanorods compacted within the films. Figure 2(b) compares the *M* vs. *T* curves of the 400°C sample for different magnetic fields, $\mu_0 H$ = 0.01, 0.1, 0.5, 1, and 2 T. No significant change in the characteristic of the magnetization was observed when the magnetic field increased, except for the slight increase of the transition temperature (Curie temperature, $T_C$). The minimum of the zero-field-cooled $dM_{ZFC}/dT$ at $\mu_0 H = 0.1$ T has been defined as the $T_C$ for the films (Fig. 2(c)). The $T_C$ values for the 300, 400 and 500 °C samples are ~275, ~277 and ~300 K, respectively. Samples with larger size of nanorods show higher $T_C$ values, which is in good agreement with work of Duong et al.[24]

Fig. 2(d) shows the Curie-Weiss fit, $1/\chi = (T - \theta)/C$, in the paramagnetic regime for the three samples. The obtained fit parameters are, $\theta = 293.650 \pm 8.516$, $296.450 \pm 4.683$, and $338.490 \pm 25.146$ K and $C = 0.244 \pm 0.005$, $0.75 \pm 0.008$, and $0.098 \pm 0.005$ emu K/mol Oe for the 300, 400, and 500 °C samples, respectively. Similarly, the effective PM moment, $\mu_{eff}$, was calculated to be $1.403 \pm 0.029$ (300 °C), $2.46 \pm 0.026$ (400 °C), and $0.9 \pm 0.046$ (500 °C) $\mu_B$. In the case of MnP, $Mn^{+3}$ exists in a low spin state, yielding a theoretical value $\mu_{eff} = 2.8\ \mu_B$[25], where $\mu_{eff} = g\sqrt{s(s+1)}\ \mu_B$. Out of the three samples, $\mu_{eff}$ of the 400 °C sample ($\mu_{eff} \sim 2.5\ \mu_B$) is closest to the theoretically calculated value ($\mu_{eff} \sim 2.8\ \mu_B$). This is associated with the highest crystallinity achieved in this sample as evident from the XRD data.

The magnetic hysteresis loops, $M$ vs. $H$, were measured at 300 K (Fig. 3(a)) and 250 K (Fig. 3(b)). As expected, the sample grown at 500 °C has the highest magnetization compared to the other two samples, since the $T_C$ of the 500 °C sample is ~300 K. It is worth noting that the $M$-$H$ loops of the 300 °C and 400 °C samples exhibit ferromagnetic behavior at room temperature, above their Curie temperatures. The FM nature can be attributed to the presence of strong magnetic anisotropy and FM correlations above $T_C$. Note that the MnP single crystal has the FM-PM transition at $T_C$ ~291 K.[26] However, depending on the growth temperature of the MnP film, nanorod sizes accompanying induced strain could alter $T_C$. Strain-modulated transition temperature changes in the MnP films grown on different substrates have also been discussed.[27, 28] Since the nanorod sizes in the 300 and 400°C samples are below 100 nm, the strain-induced effect is significant in these samples. In addition, the MnP films are polycrystalline, local anisotropies in different directions influencing the spins and their magnetic coupling could also explain the FM correlations at $T > T_C$. Fig. 3(b) shows the $M$ vs. $H$ curve at $T = 250$ K, where all the samples are in the ferromagnetic regime. The three samples show clear magnetic hysteresis loops with measurable coercivity. As expected, the highest coercivity of about 0.1 T is reported for the film grown at 500 °C (Fig. 3 (b)). As one can see clearly from the normalized $M$ vs. $H$ curves at 250 K for the films (inset of Fig. 3 (b)), the 500°C film has the strongest in-plane anisotropy. This is attributed to the presence of nanoclusters, rather than nanorods, within this film (Fig. S1). Unlike the 400 and 500°C samples, the butterfly shape of the hysteresis loop observed for the film grown at 300°C (inset of Fig. 3 (b)) implies phase coexistence; antiferromagnetic (AFM) and FM phases at low temperatures.[33] This also indicates the AFM/FM transition is strongly affected by the strain.

By subjecting the film to a change in external magnetic field at a constant temperature, the change in magnetic entropy ($\Delta S_M$) can be calculated as

$$\Delta S_M(T, \mu_0 H) = \mu_0 \int_0^{H_{max}} \left(\frac{\partial M}{\partial T}\right)_H dH. \qquad (1)$$

where $H$ is the magnetic field, $M$ is the magnetization, and $T$ is the temperature. The presence of maxima or minima in a $\Delta S_M(T)$ curve signifies the occurrence of a magnetic transition. From the magnetization vs. field $M$-$H$ data for $\mu_0 \Delta H = 0 - 5$ T, the magnetic entropy change has been calculated and summarized in Fig. 4. Figs. 4(a), (b) and (c) illustrate the $\Delta S_M(T, \mu_0 \Delta H)$ results for the 300, 400 and 500 °C samples for $\mu_0 \Delta H = 0 - 2$ T. A clear minimum at $T_C$ corresponding to the FM-PM phase transition is seen for all the samples, respectively. The calculated -$\Delta S_M$ values for $\mu_0 \Delta H = 2$ T at the transition temperatures are 0.215, 0.659, and 0.642 J/K kg for 300, 400, and 500 °C grown films, respectively. These values of -$\Delta S_M$ are lower compared to that of the MnP single crystal (~1.2 J/K kg for $\mu_0 \Delta H = 2$ T).[26] Figs. 4(d), (e) and (f) show the 2D surface plots of $\Delta S_M(T, \mu_0 \Delta H)$ for $\mu_0 \Delta H = 0 - 5$ T. The warm colors signify the lower values of -$\Delta S_M(T, \mu_0 \Delta H)$, while the cool colors represent the higher values of $-\Delta S_M(T, \mu_0\Delta H)$. From this figure, it is clear that with increase in the external magnetic field, the magnitude of $-\Delta S_M(T, \mu_0\Delta H)$ increases for all the samples, with the 400 °C sample showing the largest $-\Delta S_M$ (~0.64 and ~1.6 J/K kg for $\mu_0\Delta H$ = 2 and 5 T, respectively). Taking the specific heat value of bulk MnP ($C_P$ ~550 J/kg K)[29], it is possible to estimate an adiabatic temperature change ($\Delta T_{ad}$) of the MnP film samples around room temperature via the expression, $\Delta T_{ad} = (\Delta S_M \times T)/C_p$. We have estimated $\Delta T_{ad}$ ~ 0.3 and ~0.8 K at $\mu_0 H$ = 2 and 5 T for the 400 °C sample, respectively. In addition, the refrigerant capacity is calculated by integrating the area of the $\Delta S_M(T, \mu_0 H)$ curve. The calculated RC values are 13.09, 29.41, and 29.54 J/kg for the MnP films grown at 300, 400, and 500°C, respectively. These results indicate that the MnP films possess negligible MCE, and that the influence of the external magnetic field (needed to magnetize the MCM) on the temperature of the MnP film is insignificant.

The performance of a TE material is characterized by Seebeck coefficient ($S$) and power factor (PF), correlated through $PF=S^2/\rho$, where $\rho$ is the electrical resistivity. The PF provides a measure of conversion of heat energy into electrical energy. $S$ and $\rho$ were measured in the temperature range of 20 – 400 K.[24] Positive $S$ values indicate that the holes are the major charge carriers in the grown MnP films, which increase on raising temperature. On the other hand, $\rho$ values are of the order of $10^{-3}$ Ω cm, indicating a metallic behavior in the MnP films. It should be recalled that when the growth temperature increased, the nanorods increased in size and became closely packed (see Fig. S1(b-d)). In addition, the resistivity of the MnP film was also found to reduce with larger MnP nanorod sizes. Fig. 5(a) shows the temperature dependence of PF for all MnP films. The 300 °C sample has the smallest PF value as compared to the other two samples. The PF values of the 400 and 500 °C samples are quite close, with a sharp increase in PF at $T > 250$ K. Both samples exhibit large PF values around room temperature. We find that the 400 °C sample exhibits the highest PF.

Since the MnP film will work as a TE material in the proposed MTECD, the influence of the external magnetic field (used to magnetize the MCM) on the PF of the MnP film should also be considered.[30-32] Asamitsu et al.[30] investigated the TE effect in $La_{1-x}Sr_xMnO_3$ under magnetic field. Since $La_{1-x}Sr_xMnO_3$ ($x = 0.2$) has an insulator-metal transition at its Curie temperature (~300 K), the external magnetic field affects spin scattering which, in turn, changes $S$ and $\rho$. As a result, the sign change in $S$ was observed for this material. In pure metallic materials, however, the Mott formula accounts for the charge-carrier contribution to the Seebeck effect, the correlation between $\rho$ and $S$. The fact that the present MnP films show a metallic behavior across the entire measured temperature range of 200 – 370 K and a very small magneto-resistance ($\Delta R/R$) around room temperature ($<$ 0.7% for $\mu_0 H = 2$ T, see Fig. S2) indicates negligible influence of the applied magnetic field on the PF of the MnP films.

As pointed out above, a novel TE material in the proposed MTECD should possess combined soft magnetic, large TE and small MC properties in the room temperature range. To find an optimal cooling temperature region, we compile in Fig. 5(b) the temperature dependence of both $\Delta S_M$ ($\mu_0 \Delta H = 5$ T) and PF for the 400 and 500 °C samples. The value of PF varies from 24 – 41 µW/m K$^2$ for $T = 300 – 400$ K and its $\Delta S_M$ at $T_C = 277$ K is ~1.6 J/K kg for $\mu_0 \Delta H = 5$ T ($\Delta T_{ad}$ ~ 0.8 K at $\mu_0 H = 5$ T) for the 400 °C sample. Similarly, for the 500 °C grown film, the PF takes value from 22 – 34.6 µW/m K$^2$ for $T = 300 – 400$ K and its $\Delta S_M$ at $T_C = 300$ K is ~1.55 J/K kg for $\mu_0 \Delta H = 5$ T ($\Delta T_{ad}$ ~ 0.84 K at $\mu_0 H = 5$ T). These results indicate that in the temperature region of interest (280 – 320 K), the MnP films grown at 400 and 500 °C exhibit a ferromagnetic behavior, large TE and small MC responses. These combined properties make the MnP films are promising candidates for use in energy-efficient MTECDs. It should be noticed that while current MCE-based refrigerators operate at low frequency 5 Hz[15] due to the low heat-exchange efficiency, the proposed MTECD possesses a desirable heat flow direction and a compact structure, giving rise to the enhanced heat-exchange efficiency which allows it to operate at higher frequency (up to 50 Hz).

**4. Conclusion**

In summary, a perspective magneto-thermo-electric cooling device has been proposed utilizing a sandwich structure made of one magnetocaloric and two thermoelectric materials. The choice of the TE material that possesses a combination of magnetic softness, large TE, and small MC effects can enhance the total magnetic flux density of the MCM and the heat transfer efficiency from the MCM to the heat sink, giving rise to the enhanced cooling efficiency of the device. Films of MnP nanorods grown on Si substrates at 400 and 500°C possessing large TE, and negligible MC properties around room temperature are very promising for energy-efficient refrigeration application.

**Supporting Information**

Supporting Information is available from the Wiley Online Library or from the author.

**Acknowledgments**

Research at the University of South Florida was supported by the U.S. Department of Energy, Office of Basic Energy Sciences, Division of Materials Sciences and Engineering under Award No. DE-FG02-07ER46438. MHP, PTH and DAT acknowledge support from the VISCOSTONE USA under Grant No. 1253113200.

**Conflict of Interest**

The authors declare no conflict of interest.

**Data Availability Statement**

Research data are not shared.

**Figure Captions:**

**Fig. 1** (**a**) Sketch of a magneto-thermo-electric cooling device composed of magnetocaloric and thermoelectric materials. TEM, MCM, MCE, and K represent thermoelectric materials, magnetocaloric materials, magnetocaloric effect, and thermal conductivity, respectively; (**b**) The working principle of the proposed cooling device. Green and gray colors indicate the on and off regime respectively for the TEM, and the top/bottom part of the device is connected to the heat sink/tank.

**Fig. 2** Temperature dependence of magnetization, $M$ vs. $T$ curves, under a field-cooled-warming (FCW) protocol for (**a**) MnP films grown at 300, 400, and 500 °C at a field of 0.1 T, and for (**b**) the MnP film grown at 400°C under various magnetic fields. (**c**) the first derivative of magnetization with respect to temperature ($dM_{ZFC}/dT$) under a zero-field-cooled (ZFC) protocol and measured at 0.1 T. (**d**) Curie-Weiss fit of DC susceptibility in the paramagnetic region for the three samples.

**Fig. 3** Magnetic field dependence of magnetization, $M$ vs. $H$ curves, at (**a**) 300 K and (**b**) 250 K for MnP thin films grown on 300, 400 and 500 °C. The inset of (**b**) shows the lower field portions of the normalized $M$-$H$ loops at 250 K.

**Fig. 4** Isothermal magnetic entropy change ($\Delta S_M$) as a function of temperature for the films grown (**a**) 300 °C, (**b**) 400 °C, and (**c**) 500 °C for a field change of 2 T. The 2D-surface plots of magnetic entropy change for MnP films grown at (**d**) 300 °C, (**e**) 400 °C, and (**f**) 500 °C for field changes up to 5 T.

**Fig. 5** (**a**) The power factors (PF) of the MnP samples measured at different temperatures. (**b**) Temperature dependence of PF and $\Delta S_M$ at a field change of 5 T for the 400 and 500°C samples.

# References

1. V. Franco, J. Blázquez, J. Ipus, J. Law, L. Moreno-Ramírez and A. Conde, Progress in Materials Science **93**, 112-232 (2018).

2. J. Scott, Annual Review of Materials Research **41**, 229-240 (2011).

3. K. Engelbrecht, Journal of Physics: Energy **1** (2), 021001 (2019).

4. H. Yu, S. D. Brechet and J.-P. Ansermet, Physics Letters A **381** (9), 825-837 (2017).

5. N. De Oliveira, Journal of Applied Physics **109** (5), 053515 (2011).

6. Y. S. Ju, Journal of Electronic Packaging **132** (4) (2010).

7. A. Kitanovski, U. Plaznik, U. Tomc and A. Poredoš, International Journal of Refrigeration **57**, 288-298 (2015).

8. S. Riffat and X. Ma, International Journal of Energy Research **28** (9), 753-768 (2004).

9. L. Mañosa, A. Planes and M. Acet, Journal of Materials Chemistry A **1** (16), 4925-4936 (2013).

10. V. K. Pecharsky and K. A. Gschneidner Jr, Physical review letters **78** (23), 4494 (1997).

11. A. Barcza, M. Katter, V. Zellmann, S. Russek, S. Jacobs and C. Zimm, IEEE transactions on magnetics **47** (10), 3391-3394 (2011).

12. D. Cam Thanh, E. Brück, O. Tegus, J. Klaasse, T. Gortenmulder and K. Buschow, Journal of applied physics **99** (8), 08Q107 (2006).

13. J. Liu, T. Gottschall, K. P. Skokov, J. D. Moore and O. Gutfleisch, Nature materials **11** (7), 620-626 (2012).

14. Y. Yuan, Y. Wu, X. Tong, H. Zhang, H. Wang, X. Liu, L. Ma, H. Suo and Z. Lu, Acta Materialia **125**, 481-489 (2017).

15. A. Kitanovski and P. W. Egolf, International journal of refrigeration **33** (3), 449-464 (2010).
12

**Figure 1**

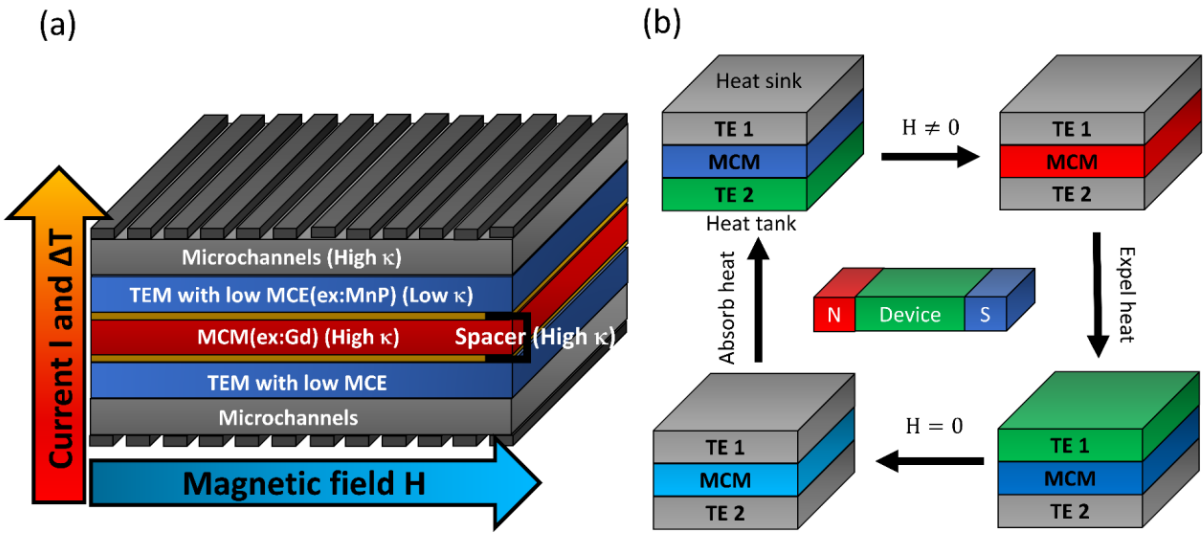

**Figure 2**

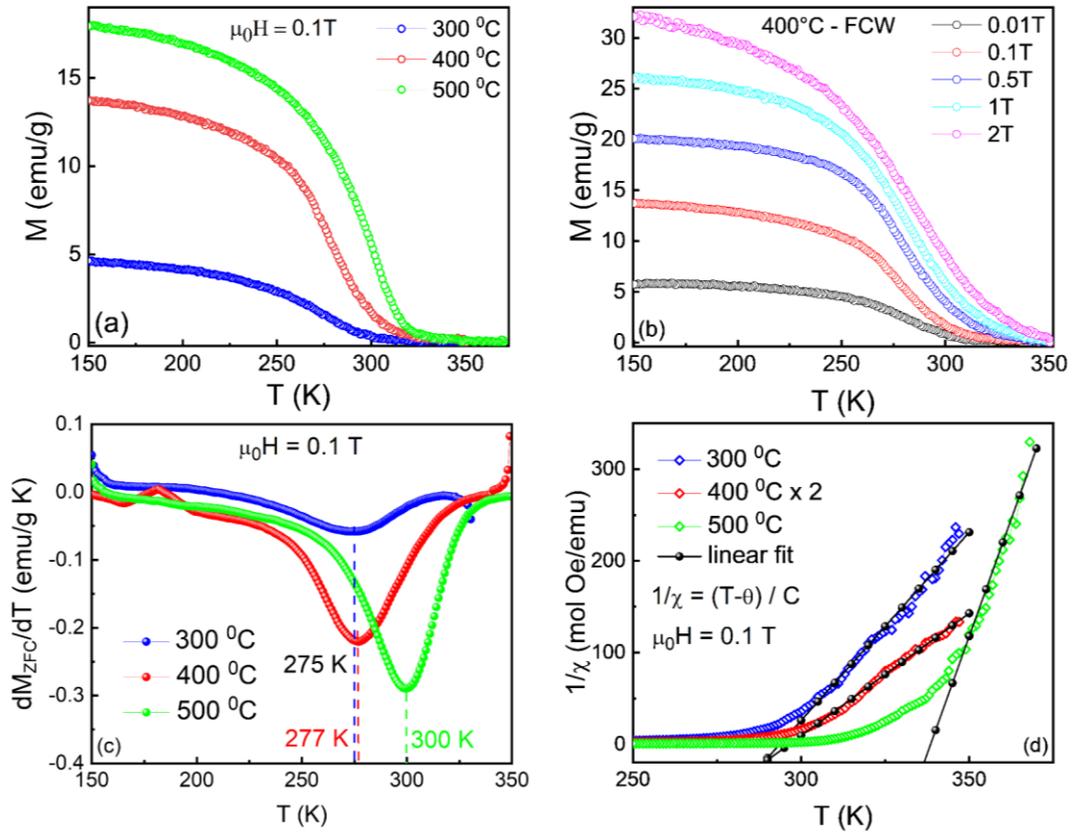

**Figure 3**

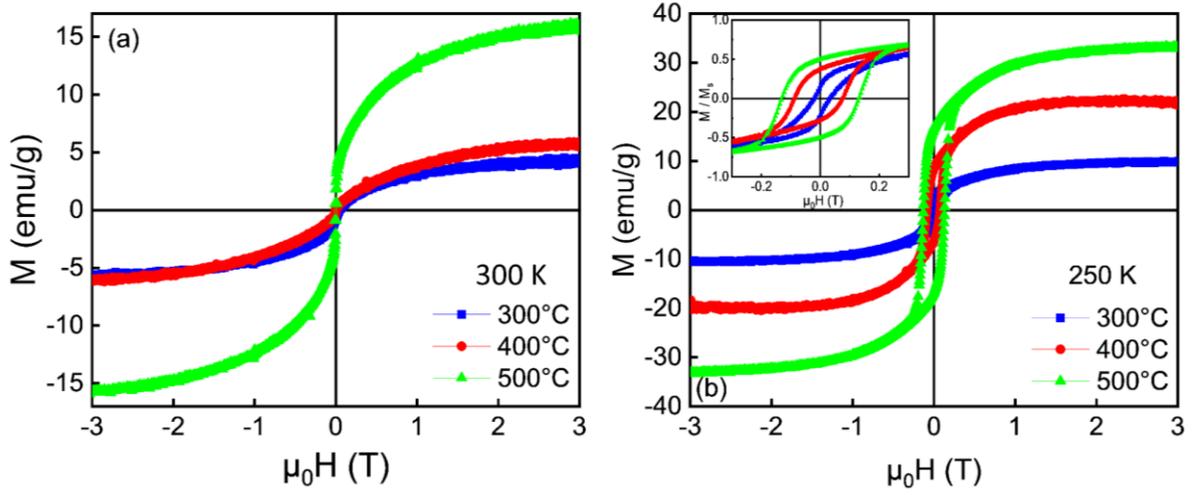

**Figure 4**

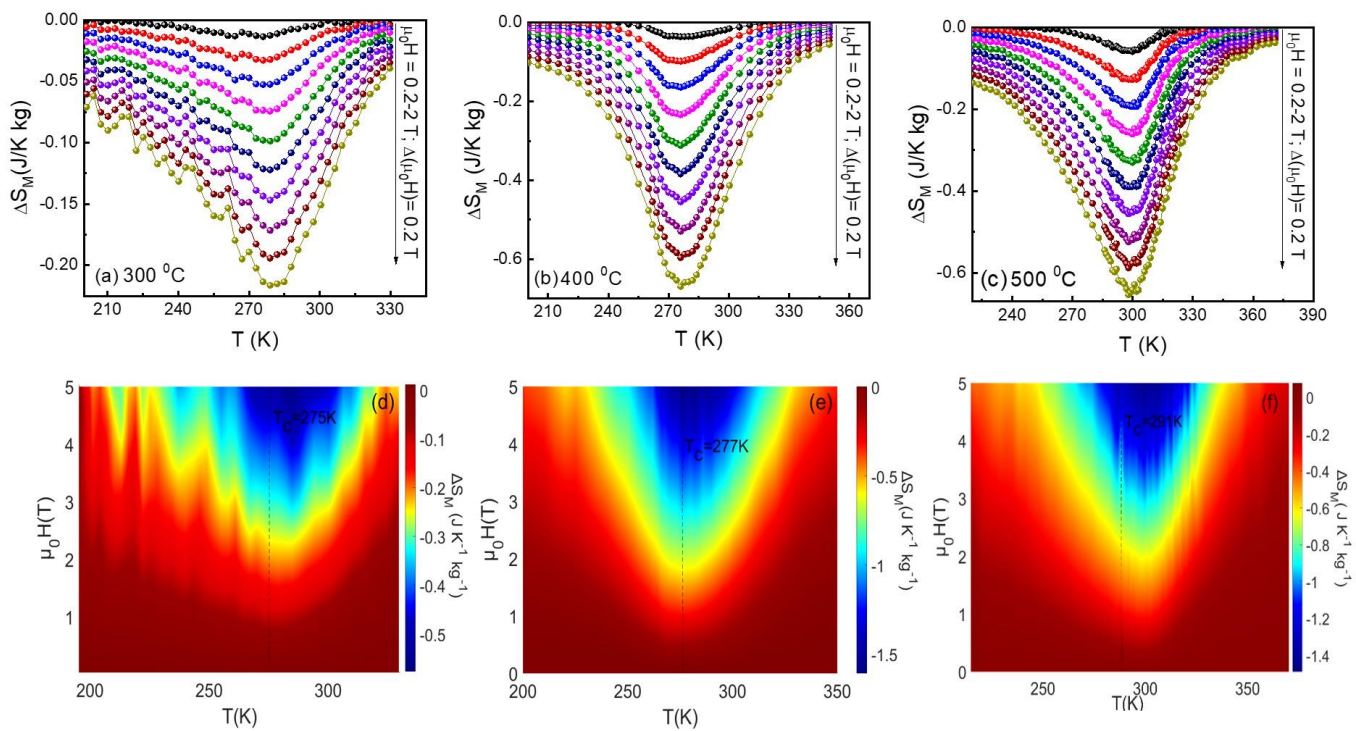

**Figure 5**

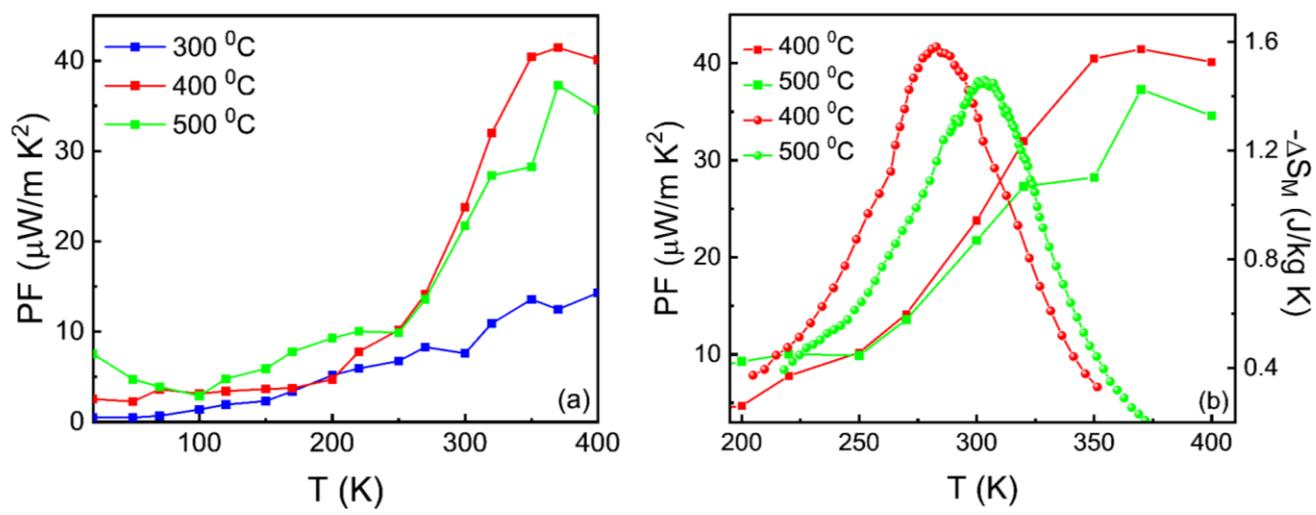
19

# Supporting Information

**MnP films with desired magnetic, magnetocaloric and thermoelectric properties for a perspective magneto-thermo-electric cooling device**

C.M. Hung[1], R.P. Madhogaria[1], B. Muchharla[1], E.M. Clements[1], A.T. Duong[2,3,*], R. Das[2], P.T. Huy[2], S.L. Cho[4], S. Witanachchi[1], H. Srikanth[1,*], and M.H. Phan[1,*]

[1] Department of Physics, University of South Florida, Tampa, Florida 33620, USA

[2] Faculty of Materials Science and Engineering, Phenikaa University, Yen Nghia, Ha Dong, Hanoi 12116, Vietnam

[3] Phenikaa Research and Technology Institute (PRATI), A&A Green Phoenix Group, 167 Hoang Ngan, Hanoi 13313, Vietnam

[4] Department of Physics and Energy Harvest-Storage Research Center, University of Ulsan, Ulsan 680-749, Republic of Korea

*Corresponding authors: tuan.duonganh@phenikaa-uni.edu.vn (A.T.D); sharihar@usf.edu (H.S); phanm@usf.edu (M.H.P)

**Figure S1**

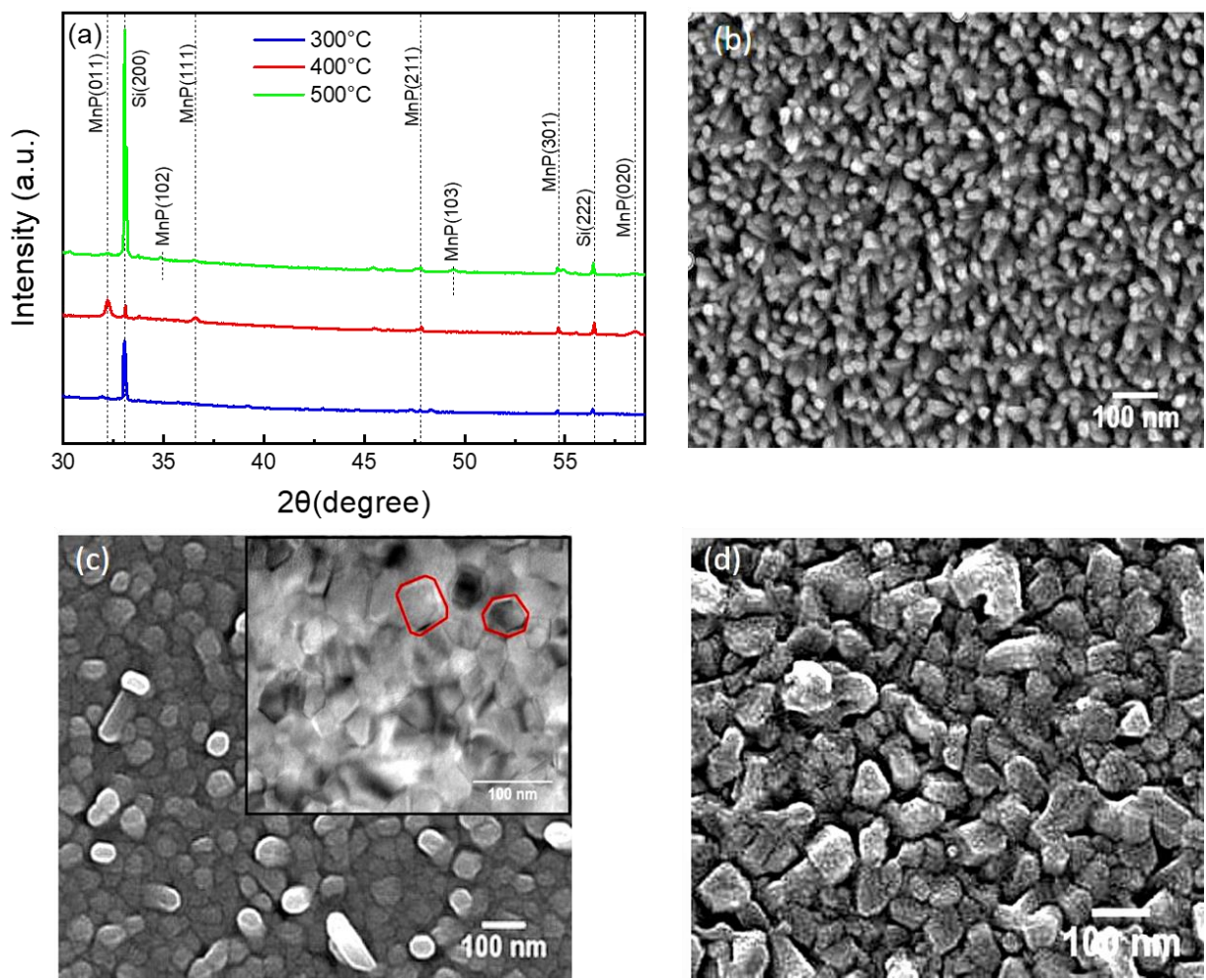

**Fig. S1** Structural characterization of the MnP films: (**a**) XRD patterns and SEM images of MnP films grown at (**b**) 300°C, (**c**) 400 °C, and (**d**) 500 °C. Inset of (**c**) is a top-view TEM image showing MnP nanorods.

**Figure S2**

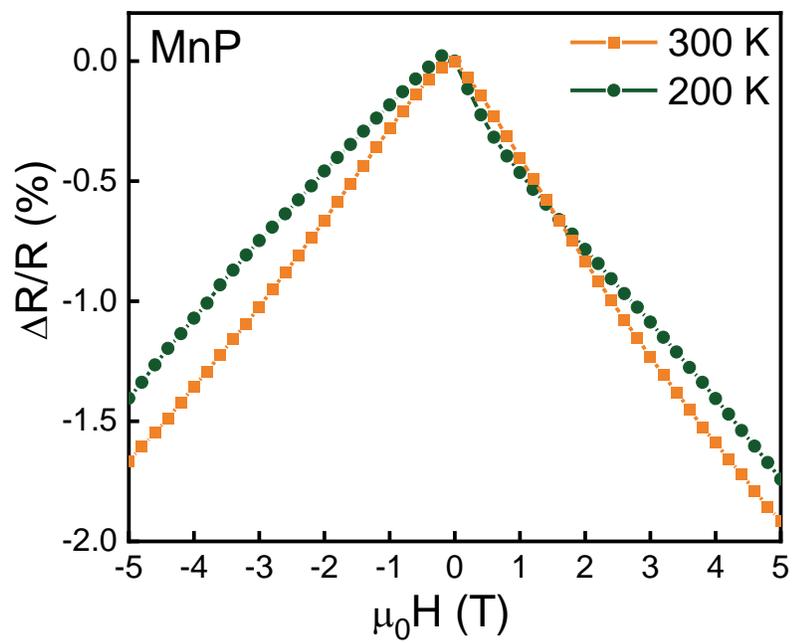

**Fig. S2.** Magnetoresistance of the 500 °C sample taken at 200 and 300 K.